\documentclass[sigconf,screen]{acmart}

\usepackage{microtype}

\usepackage{multirow}
\usepackage{tabularx}
\usepackage{cleveref}
\usepackage{multirow}
\usepackage{csvsimple}

\crefname{section}{\S}{\S\S}
\Crefname{section}{\S}{\S\S}

\AtBeginDocument{\providecommand\BibTeX{{\normalfont B\kern-0.5em{\scshape i\kern-0.25em b}\kern-0.8em\TeX}}}

\setcopyright{acmlicensed}
\acmPrice{15.00}
\acmDOI{10.1145/3468264.3473113}
\acmYear{2021}
\copyrightyear{2021}
\acmSubmissionID{fse21demo-p7-p}
\acmISBN{978-1-4503-8562-6/21/08}
\acmConference[ESEC/FSE '21]{Proceedings of the 29th ACM Joint European Software Engineering Conference and Symposium on the Foundations of Software Engineering}{August 23--28, 2021}{Athens, Greece}
\acmBooktitle{Proceedings of the 29th ACM Joint European Software Engineering Conference and Symposium on the Foundations of Software Engineering (ESEC/FSE '21), August 23--28, 2021, Athens, Greece}

\renewcommand{\comment}[1]{}

\renewcommand{\subsubsection}[1]{\noindent \textbf{#1:}\ }

\begin{document}

\graphicspath{{graphics/}}

\title[Sangrahaka: A Tool for Annotating and Querying Knowledge Graphs]{Sangrahaka: A Tool for Annotating and Querying\\ Knowledge Graphs}

\author{Hrishikesh Terdalkar}
\email{hrishirt@cse.iitk.ac.in}
\orcid{0000-0001-8343-424X}
\affiliation{
  \institution{Indian Institute of Technology Kanpur}
  \country{India}
}

\author{Arnab Bhattacharya}
\email{arnabb@cse.iitk.ac.in}
\orcid{0000-0001-7331-0788}
\affiliation{
  \institution{Indian Institute of Technology Kanpur}
  \country{India}
}

\begin{abstract}
We present a web-based tool \emph{Sangrahaka} for annotating entities and
	relationships from text corpora towards construction of a knowledge graph
	and subsequent querying using templatized natural language questions.  The
	application is language and corpus agnostic, but can be tuned for specific
	needs of a language or a corpus.  The application is freely available for
	download and installation.  Besides having a user-friendly interface, it is
	fast, supports customization, and is fault tolerant on both client and
	server side.  It outperforms other annotation tools in an objective
	evaluation metric.  The framework has been successfully used in two
	annotation tasks.  The code is available from
	\emph{\url{https://github.com/hrishikeshrt/sangrahaka}}.
\end{abstract}



\maketitle

\section{Introduction and Motivation}
\label{sec:intro}

\emph{Annotation} is a process of marking, highlighting or extracting relevant
information from a corpus.  It is important in various fields of computer
science including natural language processing (NLP) and text mining.  A generic
application of an annotation procedure is in creating a dataset that can be
used as a training or testing set for various machine learning tasks. The exact
nature of the annotation process can vary widely based on the targeted task,
though.

In the context of NLP, annotation often refers to identifying and highlighting
various parts of the sentence (e.g., characters, words or phrases) along with
syntactic or semantic information.  Semantic tasks are high-level tasks dealing
with the meaning of linguistic units and are considered among the most
difficult tasks in NLP for any language. {\em Question Answering (QA)} is an
example of a semantic task that deals with answering a question asked in a
natural language. The task requires a machine to `understand' the language,
i.e., identify the intent of the question, and then search for relevant
information in the available text.  This often encompasses other NLP tasks such
as parts-of-speech tagging, named entity recognition, co-reference resolution,
and dependency parsing \cite{jurafsky2000speech}.

Making use of knowledge bases is a common approach for the QA task
\cite{voorhees1999trec, hirschman2001natural, kiyota2002dialog,
yih2015semantic}.  Construction of knowledge graphs (KGs) from free-form text,
however, can be very challenging, even for English.  The situation for other
languages, whose state-of-the-art in NLP is not as advanced in English, is
worse.  As an example, consider the epic
\emph{Mahabharata}\footnote{Mahabharata is one of the two epics in India (the
other being Ramayana) and is probably the largest book in any literature,
containing nearly 1,00,000 sentences. It was originally composed in Sanskrit.}
in Sanskrit.  The state-of-the-art in Sanskrit NLP is, unfortunately, not
advanced enough to identify the entities in the text and their
inter-relationships.  Thus, human annotation is currently the only way of
constructing a KG from it.

Even a literal sentence-to-sentence translation of Mahabharata in English,
which probably boasts of the best state-of-the-art in NLP, is not good enough.
Consider, for example, the following sentence from (an English translation of)
``The Mahabharata'' \cite{ganguli1884mahabharata}:

\begin{quotation}
{\em Ugrasrava}, the son of {\em Lomaharshana}, surnamed {\em Sauti},
	well-versed in the {\em Puranas}, bending with humility, one day approached
	the great sages of rigid vows, sitting at their ease, who had attended the
	twelve years' sacrifice of {\em Saunaka}, surnamed {\em Kulapati}, in the
	forest of {\em Naimisha}.
\end{quotation}

The above sentence contains numerous entities, e.g., {\em Ugrasrava}, {\em
Lomaharshana}, as well as multiple relationships, e.g., {\em Ugrasrava
is-son-of Lomaharshana}.  One of the required tasks in building a KG for
Mahabharata is to extract these entities and relationships.

Even a state-of-the-art tool such as {\em spaCy}~\cite{spacy} makes numerous
mistakes in identifying the entities; it misses out on {\em Ugrasrava} and
identifies types wrongly of several entities, e.g., {\em Lomaharshana} is
identified as an {\em Organization} instead of a {\em Person}, and {\em
Saunaka} as a {\em Location} instead of a {\em Person}.\footnote{This is not a
criticism of spaCy; rather, this highlights the hardness of semantic tasks such
as entity recognition.} Consequently, relationships identified are also
erroneous.  This highlights the difficulty of the task for machines and
substantiates the need for human annotation.

\section{Background}
\label{sec:background}

There are numerous text annotation tools available including {\em WebAnno}
\cite{yimam2013webanno}, {\em FLAT} \cite{flat2014}, {\em BRAT}
\cite{stenetorp2012brat}, {\em GATE Teamware} \cite{bontcheva2013gate}, and
{\em doccano} \cite{doccano}, for handling a variety of text annotation tasks
including classification, labeling and sequence-to-sequence annotations.

Ideally, an annotation tool should support multiple features and facilities
such as friendly user-interface, simple setup and annotation, fairly fast
response time, distributed framework, web-based deployment, customizability,
access management, crash tolerance, etc.  In addition, for the purpose of
knowledge-graph focused annotation, it is important to have capabilities for
multi-label annotations and support for annotating relationships.

While {\em WebAnno} is extremely feature rich, it compromises on simplicity.
Further, its performance deteriorates severely as the number of lines displayed
on the screen increases.  {\em GATE} also has the issue of complex installation
procedure and dependencies. {\em FLAT} has a non-intuitive interface and
non-standard data format. Development of {\em BRAT} has been stagnant, with the
latest version being published as far back as 2012. The tool {\em doccano},
while simple to setup and use, does not support relationship annotation.  Thus,
unfortunately, none of these tools supports all the desired features of an
annotation framework for the purpose of knowledge-graph annotation. Further,
none of the above frameworks provide an integration with a graph database, a
querying interface, or server and client side crash tolerance.

Thus, to satisfy the need of an annotation tool devoid of these pitfalls, we
present \emph{Sangrahaka}.  It allows users to annotate and query through a
single platform. The application is language and corpus agnostic, but can be
customized for specific needs of a language or a corpus.
Table~\ref{table:annocompare} provides a high-level feature comparison of these
annotation tools including \emph{Sangrahaka}.

\begin{table}[t]
    \caption{Comparison of features of various annotation tools}
    \vspace*{-4mm}
    \resizebox{\columnwidth}{!}{
    \begin{tabular}{ccccccc}
        \toprule
        \bf Feature & WebAnno & GATE & BRAT & FLAT & doccano & Sangrahaka\\
        \midrule
        Distributed Annotation & \checkmark & \checkmark & \checkmark & \checkmark & \checkmark & \checkmark\\
        Simple Installation & ~ & ~ & \checkmark & \checkmark & \checkmark & \checkmark\\
        Intuitive & \checkmark & \checkmark & \checkmark & ~ & \checkmark & \checkmark\\
        Entity and Relationship & \checkmark & \checkmark & \checkmark & \checkmark & ~ & \checkmark\\
        Query Support & ~ & ~ & ~ & ~ & ~ & \checkmark\\
        Crash Tolerance & ~ & ~ & ~ & ~ & ~ & \checkmark\\
        \bottomrule
    \end{tabular}
    }
    \vspace*{-4mm}
	\label{table:annocompare}
\end{table}

A recently conducted extensive survey \cite{neves2021extensive} evaluates $78$
annotation tools and provides an in-depth comparison of $15$ tools. It also
proposes a scoring mechanism by considering $26$ criteria covering publication,
technical, function and data related aspects.  We evaluate \emph{Sangrahaka}
and other tools using a the same scoring mechanism, albeit with a modified set
of criteria.  The details are in~\cref{sec:evaluation}.

\section{Architecture}
\label{sec:tool}

\emph{Sangrahaka} is a language and corpus agnostic tool. Salient features of
the tool include an interface for annotation of entities and relationships, and
an interface for querying using templatized natural language questions. The
results are obtained by querying a graph database and are depicted in both
graphical and tabular formats.  The tool is also equipped with an
administrators' interface for managing user access levels, uploading corpora
and ontology creation. There are utility scripts for language-specific or
corpus-specific needs. The tool can be deployed on the Web for distributed
annotation by multiple annotators.  No programming knowledge is expected from
an annotator.

The primary tools and technologies used are {\em Python 3.8} \cite{python3},
{\em Flask 1.1.2} \cite{ronacher2011flask, grinberg2018flask}, {\em Neo4j
Community Server 4.2.1} \cite{webber2012neo4j}, {\em SQLite 3.35.4}
\cite{sqlite2021hipp} for the backend and {\em HTML5}, {\em JavaScript}, {\em
Bootstrap 4.6} \cite{bootstrap46}, and {\em vis.js} \cite{visjs} for the
frontend.

\subsection{Workflow}

\begin{figure*}[t]
    \vspace*{-2mm}
    \includegraphics[width=0.75\textwidth]{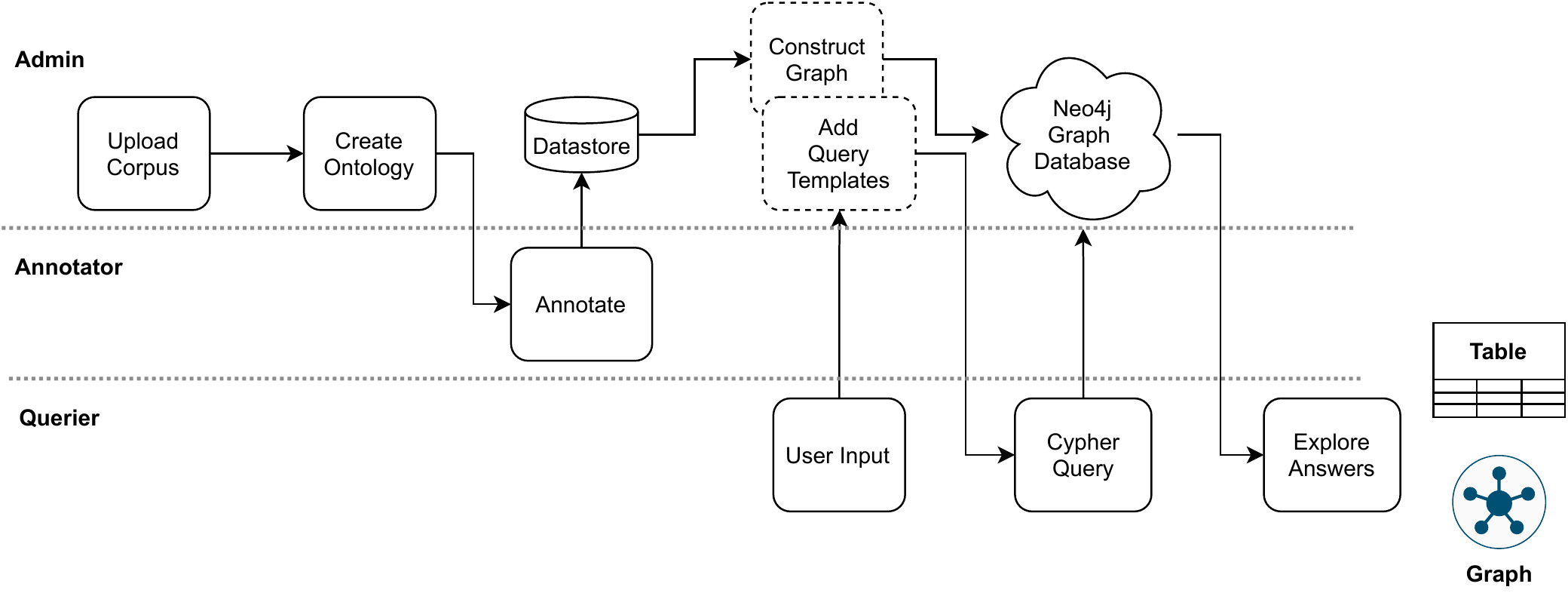}
    \vspace*{-6mm}
    \caption{Architecture and Workflow}
    \Description{Workflow of Admin, Annotator and Querier roles and their interaction with each other.
    Corpus creation, ontology creation, annotation, graph creation, graph querying are the principal components.}
    \label{fig:architecture}
    \vspace*{-2mm}
\end{figure*}

Figure~\ref{fig:architecture} shows the architecture and workflow of the
system.

\begin{table}[t]
    \caption{Roles and Permissions}
	\vspace*{-4mm}
    \resizebox{0.75\columnwidth}{!}
	{
	    \begin{tabular}{ccccc}
	        \toprule
	        \multirow{2}{*}{\bf Permissions} & \multicolumn{4}{c}{\bf Roles}\\
	        \cline{2-5}
	        & Querier & Annotator & Curator & Admin\\
	        \midrule
	        Query & \checkmark & \checkmark & \checkmark & \checkmark\\
Annotate & ~ & \checkmark & \checkmark & \checkmark\\
Curate & ~ & ~ & \checkmark & \checkmark\\
Create Ontology & ~ & ~ & ~ & \checkmark\\
Upload Corpus & ~ & ~ & ~ & \checkmark\\
Manage Access & ~ & ~ & ~ & \checkmark\\
	        \bottomrule
	    \end{tabular}
	}
	\label{table:access}
	\vspace*{-4mm}
\end{table}

The tool is presented as a web-based full-stack application.  To deploy it, one
first configures the application and starts the server.  A user can then
register and login to access the interface. The tool uses a role based access
system. Roles are Admin, Curator, Annotator, and Querier.  Permissions are tied
to roles.  Table~\ref{table:access} enlists the roles and the permissions
associated with them. A user can have more than one role.  Every registered
member has permission to access user control panel and view corpus.

An administrator creates a corpus by uploading the text. She also creates a
relevant ontology for the corpus and grants annotator access to relevant users.
The ontology specifies the type of entities and relationships allowed.  An
annotator signs-in and opens the corpus viewer interface to navigate through
lines in the corpus.  For every line, an annotator then marks the relevant
entities and relationships.  A curator can access annotations by all
annotators, and can make a decision of whether to keep or discard a specific
annotation. This is useful to resolve conflicting annotations.  An
administrator may customize the graph generation mechanism based on the
semantic task and semantics of the ontology. She then imports the generated
graph into an independently running graph database server.  A querier can then
access the querying interface and use templatized natural language questions to
generate graph database queries. Results are presented both in graphical as
well as tabular formats and can be downloaded as well.

\subsection{Backend}

The backend is written in {\em Python}, using {\em Flask}, a microwebframework.
Pluggable components of the backend are a relational database and a Neo4j
graph database server.

\subsubsection{Web Framework}
The web framework manages routing, templating, user-session management,
connections to databases, and other backend tasks. A {\em Web Server Gateway
Interface (WSGI)} HTTP server runs the Flask application.
We use {\em Gunicorn} \cite{gunicorn} running behind an {\em NGINX}
\cite{nginx} reverse proxy for this purpose.  However any WSGI server,
including the Flask's in-built server, can be used.

\subsubsection{Data}
Data related to user accounts, roles as well as corpus text, ontology, entity
annotations and relationship annotations are stored in a relational database.
This choice is made due to the need of cross-references (in database parlance,
\emph{joins}) across user, corpus and annotation related information.  Any
relational database compatible with {\em SQLAlchemy} \cite{sqlalchemy} can be
used.  We have used {\em SQLite}. An administrator uploads various chapters in
a corpus as JSON files using a pre-defined format. Each JSON object contains
text of the line and optional extra information such as word segmentation,
verse id, linguistic information etc.  The structure of JSON file corresponding
to a chapter is explained in Appendix~A of \cite{sangrahakaarxiv}. This
information is then organized in a hierarchical structure with 4 levels: {\em
Corpus}, {\em Chapter}, {\em Verse} and {\em Line}.  Additionally, there is an
{\em Analysis} table that stores the linguistic information for each line.  The
system is equipped to deal with morphologically rich languages. Lemmas (i.e.,
word roots) are stored in a separate table and referenced in entity and
relationship annotations.  Every entity annotation consists of a lemma, an
entity type, a line number and user-id of the annotator.  Every relationship
annotation consists of a source (lemma), a target (lemma), a relationship type,
an optional detail text, a line number and user-id of the annotator.

\subsubsection{Knowledge Graph}
A knowledge graph is constructed using the entity and relationship annotations.
{\em Neo4j} is used as the graph database server to store and query the KG.
Connection to it is made using the {\em Bolt} protocol \cite{boltprotocol}.
Hence, the graph database can exist independently on a separate system.  {\em
Cypher} query language \cite{cypher} is used to query the graph database and
produce results.

\subsubsection{Natural Language Query Templates}
Templates for natural language questions are added by an administrator. A query
template has two essential components, a natural language question with
placeholder variables and a {\em Cypher} equivalent of the query with
references to the same placeholder variables.  Placeholder variables represent
values where user input is expected. Query templates are provided in a JSON
file whose structure is explained in Appendix~A of \cite{sangrahakaarxiv}.  The
natural language query template, combined with user input, forms a valid
natural language question, and the same replacement in {\em Cypher} query
template forms a valid {\em Cypher} query.

\subsubsection{Configuration}
The application contains several configurable components. The entire configuration setting
is stored in a settings file.  Table~\ref{table:config} explains some important
configuration options.

\begin{table}[t]
    \caption{Main configuration options}\label{table:config}
    \vspace*{-4mm}
	\resizebox{\columnwidth}{!}
	{
	{\small
    \begin{tabular}{p{0.25\columnwidth}p{0.70\columnwidth}}
        \toprule
        {\bf Option} & {\bf Explanation} \\
        \midrule
        Admin user & Username, Password and E-mail of owner \\
        Roles & Configuration of Roles and Permissions \\
SQL config & {\em SQLAlchemy} compatible Database URI \\
        Neo4j config & Server URL and Credentials \\
\bottomrule
    \end{tabular}
	}
	}
    \Description{List of important configuration options and their explanation}
    \vspace*{-4mm}
\end{table}

\begin{figure*}[t]
	\vspace*{-2mm}
    \includegraphics[width=0.50\textwidth]{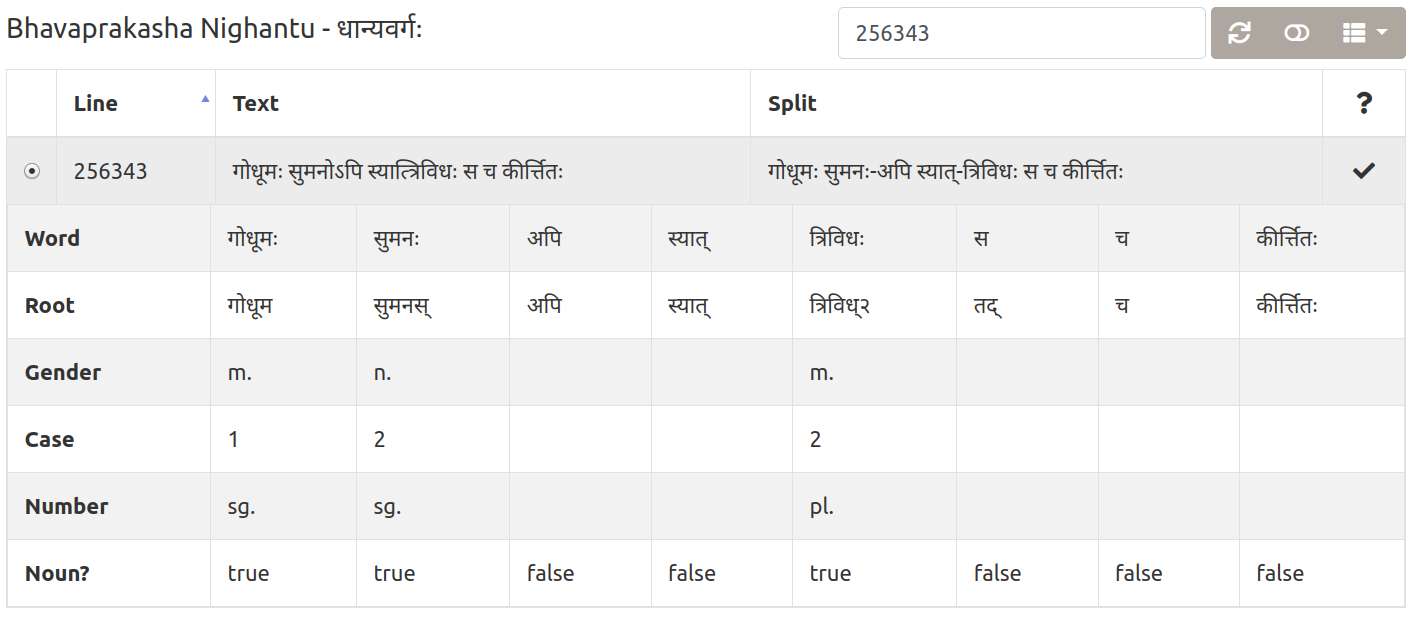}
    \includegraphics[width=0.25\textwidth]{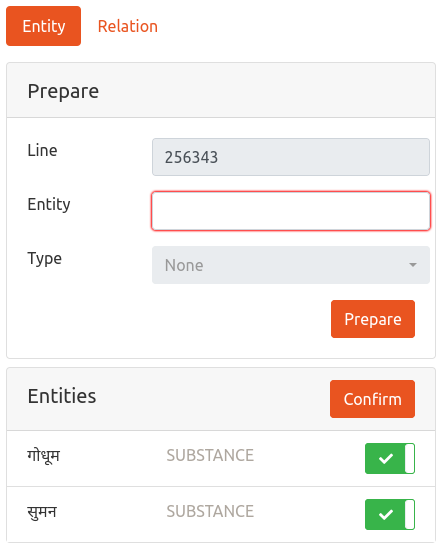}
    \includegraphics[width=0.24\textwidth]{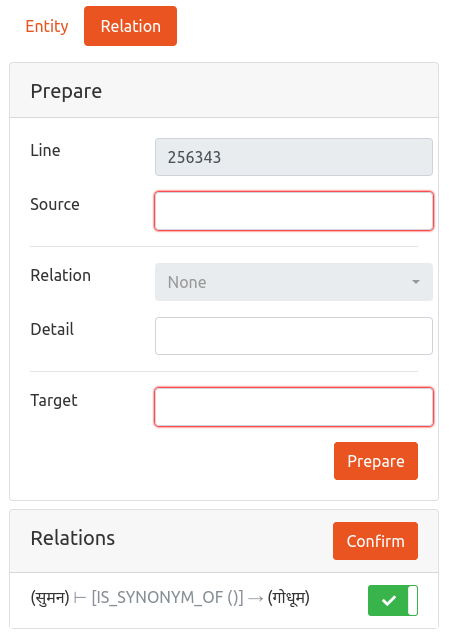}
    \includegraphics[width=0.50\textwidth]{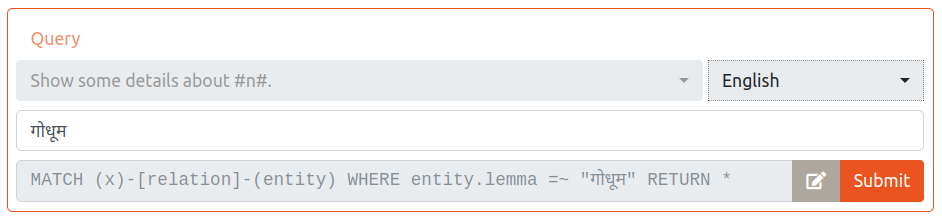}
    \includegraphics[width=0.20\textwidth]{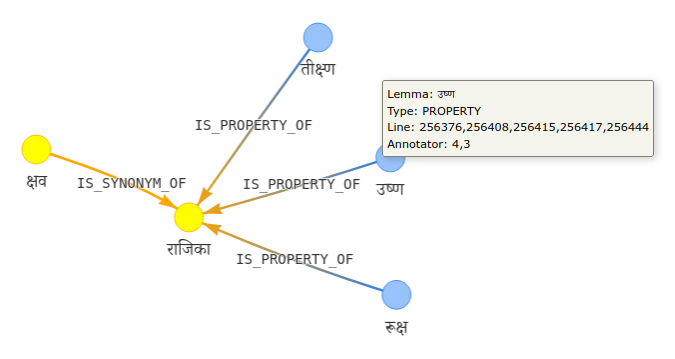}
    \includegraphics[width=0.29\textwidth]{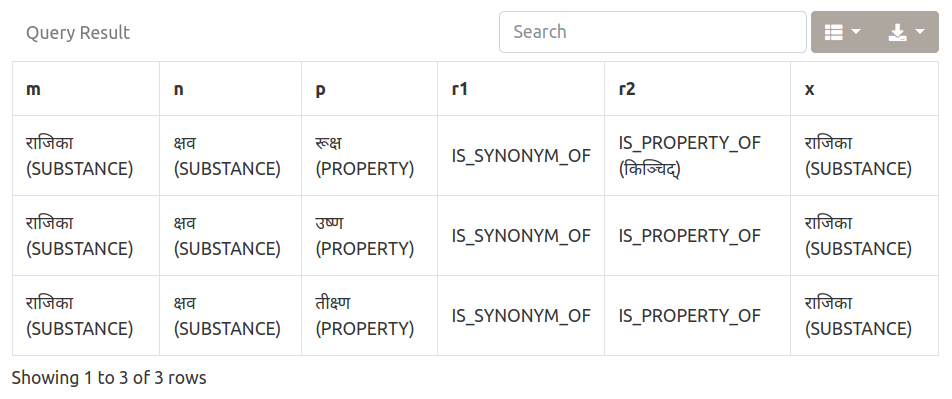}
	\vspace*{-6mm}
    \caption{Interfaces: Corpus Viewer, Entity Annotator, Relation Annotator,
             Query Interface, Graphical Result, Tabular Result}
    \Description{Corpus Viewer, Entity Annotator, Relation Annotator,
                 Query Interface, Graphical Result Interface,
                 Tabular Result Interface}
	\vspace*{-2mm}
    \label{fig:interface}
\end{figure*}

\subsubsection{Utility Scripts}
Utility Python scripts are provided for tasks that need to be performed in the
background.  The primary among these is a graph generation script to generate
JSONL \cite{jsonl} formatted data suitable for direct import in the {\em Neo4j
Graph Database}.  Sample scripts are also provided for generation of corpus
file and query template file. These can be easily customized to suit corpus
specific or application specific needs.

\subsection{Frontend}

The frontend is in form of a web application. HTML5 webpages are generated
using {\em Jinja} template engine \cite{jinja2}, styled using {\em Bootstrap
4.6} and made interactive using {\em JavaScript}.  The web-based user interface
has several components that are accessible to users based on their roles.  The
user interface is shown in Figure~\ref{fig:interface}.

\subsubsection{Corpus Viewer Interface}
The corpus viewer interface consists of a row-wise display of lines in a
corpus.  For such languages such as Sanskrit, German, Finnish, Russian, etc.
that exhibit a large number of compound words, the corpus viewer can display
the word-split output added by the administrator.  Further, an
administrator may run other language specific tools to obtain any kind of
semantic and syntactic information about the components of the sentence as a
list of key-value pairs.  The corpus viewer displays this information in a
tabular format whenever a line is selected.

\subsubsection{Annotator Interface}
The annotator interface is interlinked with the corpus viewer interface.  It
contains two views, one for entity annotation and the other for relation
annotation.  Adaptive auto-complete suggestions are offered based on previously
added lemmas and lemmas present in the line being annotated.

\subsubsection{Query Interface}
The query interface makes use of pre-defined natural language query templates
and combines them with user input to form {\em Cypher} queries. A user may
directly edit the {\em Cypher} query as well if she so desires.  These are
communicated to the graph database using {\em Bolt} protocol and results are
fetched.  Result of a {\em Cypher} query is a subgraph of the knowledge graph
and is presented in an interactive interface that allows users to zoom-in to
specific areas of the graph, rearrange nodes and save the snapshot of the graph
as an image.  Results are also displayed in a tabular manner and can be
exported in various file formats including CSV, JSON, text, etc.

\subsubsection{Admin Interface}
The administrator frontend allows an administrator to perform tasks such as
change users' access levels, create corpus, upload chapters in a corpus, and
create ontology.  Adding a new corpus requires two steps: corpus creation and
chapter upload. The corpus creation step refers to creating a new entry in the
{\em Corpus} table along with a description. Once a corpus has been added,
chapters associated with the corpus can be uploaded.

\subsubsection{Ontology Creation}
The ontology creation interface allows an administrator to add or remove node
types and relation types.  If an entity or relation type is being used in an
annotation, removal of the same is prevented.

\subsubsection{Curation}
Curation is performed through the annotation interface. A curator can see
annotations by all annotators and can choose to keep or remove them as well as
add new annotations.

\begin{table}[t]
	\caption{Annotation tasks performed using {\em Sangrahaka}}
    \vspace*{-4mm}
    \resizebox{\columnwidth}{!}{
    \begin{tabular}{rrrrrrrr}
        \toprule
		\multirow{2}{*}{\bf Corpus} & \multirow{2}{*}{\bf Lines} & \multirow{2}{*}{\bf Annotators} & \multicolumn{2}{c}{\bf Ontology} & \multicolumn{2}{c}{\bf Annotations} & \multirow{2}{*}{\bf Progress} \\
        ~ & ~ & & Nodes & Relations & Nodes & Relations \\
        \midrule
        BPN \cite{bpn2016} & $180$ & $5$ & $25$ & $30$ & $602$ & $778$ & $100\%$ \\
        VR \cite{valmikiramayana} & $17655$ & $9$ & $107$ & $132$ & $1810$ & $2087$ & $54\%$ \\
        \bottomrule
    \end{tabular}
    }
	\label{table:annotationtasks}
    \vspace*{-4mm}
\end{table}

\subsection{Fault Tolerance}

Corpus viewer and annotation interface act as a {\em single-page application}
\cite{wikispa} and make use of {\em AJAX} \cite{wikiajax} calls to communicate
with the server.  Entity and Relation annotation processes have two steps,
`Prepare' and `Confirm'.  Once an entity or a relation is prepared, it is
stored in a browser based {\tt localStorage} \cite{Hickson:21:WSE} that
persists across browser sessions and is, thus, preserved even if the browser
crashes.  Once a user clicks `Confirm', an attempt to contact the server is
made. If the attempt is successful and the data is inserted in the database
successfully, the server returns success and the data associated with that
annotation is cleared from the local storage. If the server returns failure,
the data persists. Thus, a server crash does not affect the user's unconfirmed
annotations. Further, if a page is already loaded in the browser and the server
crashes, a user can still continue to annotate. The unconfirmed entities and
relations are color coded and can be easily located later to confirm once the
server is restored.  Thus, the application is fault-tolerant on both client and
server side.  This is an important feature that distinguishes \emph{Sangrahaka}
from other tools.

\section{Evaluation}
\label{sec:evaluation}

The tool has been used for two distinct annotation tasks: (1)~a chapter from a
medical text (\emph{Ayurveda}) corpus in Sanskrit ({\em BPN}) \cite{bpn2016},
and (2)~full text of the epic \emph{Ramayana} in English ({\em VR})
\cite{valmikiramayana}.  Table~\ref{table:annotationtasks} presents details of
these tasks.

As a subjective evaluation of the tool, the annotators from both the tasks,
were asked to rate the tool on a scale of 5. We received an overall rating of
$4.5$ from $10$ annotators.

As an objective evaluation, we followed the methodology used in
\cite{neves2021extensive}.  We dropped parameters linked to publication and
citations and those that were not met by any tool.  Instead, we added 4 new
criteria, namely (1)~support for querying, (2)~server side and (3)~client side
crash tolerance and (4)~distributed annotation support. We re-evaluated the
three highest-scoring tools ({\em WebAnno}, {\em BRAT}, and {\em FLAT}) from
\cite{neves2021extensive} as well as \emph{Sangrahaka} using the modified set
of $25$ criteria. {\em Sangrahaka} outperformed other tools with a score of
$0.82$ compared to {\em FLAT} ($0.78$), {\em WebAnno} ($0.74$) and {\em BRAT}
($0.70$).
Further details about the evaluation can be found in Appendix~B of
\cite{sangrahakaarxiv}.

\section{Conclusions and Future Work}
\label{sec:conclusion}

In this paper, we have proposed a web-based tool \emph{Sangrahaka} for
annotation and querying of knowledge graphs.

In future, we plan to streamline the process of running utility scripts and
third-party NLP tools directly through the frontend. We also aim to explore
ways of resolving conflicting annotations automatically. Further, we will
release more user-friendly installation options such as Docker, VM, etc.

\pagebreak

\bibliographystyle{ACM-Reference-Format}
\bibliography{papers}

\pagebreak

\appendix
\section{Data Examples}
\label{sec:dataexamples}

The tool primarily uses JSON data format for various purposes such as corpus
input and query template definitions.

\subsection{Corpus Format}

The outermost structure is a list containing objects corresponding to lines.
The keys of a line object include \emph{text} (line text), \emph{split}
(optional text with word segmentation), \emph{verse} (optional verse id in case
of poetry), \emph{analysis} (optional linguistic information as a list of
key-value pairs, corresponding to tokens in the sentence).

Following is an example of a corpus file containing a single sentence.

\begin{verbatim}
[
    {
        "verse": 1,
        "text": "To sainted Nárad, prince of those",
        "split": "",
        "analysis": {
        "source": "spacy",
        "text": "",
        "tokens": [
            {
                "Word": "Nárad",
                "Lemma": "Nárad",
                "Tag": "NNP",
                "POS": "PROPN",
            },
            {
                "Word": "prince",
                "Lemma": "prince",
                "Tag": "NN",
                "POS": "NOUN",
            }
        ]
    },
]
\end{verbatim}

\subsection{Query Template}

The outermost structure is a list, containing query objects.  Each query object
corresponds to a single query template and has the keys \emph{gid},
\emph{groups}, \emph{texts}, \emph{cypher}, \emph{input} and \emph{output}
where \emph{gid} is used to group similar queries together in the frontend, and
\emph{groups} and \emph{texts} are objects that contain language names as keys,
and names of groups and query texts in those languages as values. If a query
expects user input, those are specified as {\tt \{0\}}, {\tt \{1\}} in the
query text. The value of key \emph{input} is a list of objects containing
information to populate frontend user-input elements.  Every object should have
a unique \emph{id} for the element and a \emph{type} of the input element. The
valid types are \emph{entity}, \emph{entity\_type}, \emph{relation} and
\emph{relation\_detail}.

\pagebreak

Following is an example of a query template file containing a single query,

\begin{verbatim}
[
    {
        "gid": "1",
        "cypher": "MATCH (p1)-[r:IS_FATHER_OF]->(p2)
                   WHERE p2.lemma =~ \"{0}\" RETURN *",
        "input": [
            {
                "id": "p",
                "type": "entity"
            }
        ],
        "output": ["p1", "r", "p2"],
        "texts": {
            "english": "Who is the father of {0}?",
        },
        "groups": {
            "english": "Kinship",
        }
    },
]
\end{verbatim}

\section{Evaluation}
\label{sec:details}

Due to the nature of semantic annotation, where an annotator usually has to
spend more time on mentally processing the text to decide the entities and
relationships than the actual mechanical process of annotating the text, and
the fact that annotations are done over several sessions of various lengths
over an extended period of time, time taken for annotation is not an adequate
metric of evaluation.

\subsection{Subjective Evaluation}

\begin{table}[t]
    \begin{tabular}{cc}
        \toprule
        Metric & Score\\
        \midrule
        Looks and feel & $4.7$\\
        Ease of use & $4.4$\\
        Annotation Interface & $4.5$\\
        Querying Interface & $4.6$\\
        Administrative Interface & $4.7$\\
        \midrule
        Overall	& $4.5$\\
        \bottomrule
    \end{tabular}
    \caption{User ratings}
	\label{table:surveyratings}
\end{table}

As a subjective evaluation, a survey was conducted among the annotators from
two annotation tasks. They were asked to rate the tool on a scale of 5 in
several metrics. The survey also asked them to describe their experience with
{\em Sangrahaka}. A total of $10$ annotators participated in the survey.

\begin{figure}[t]
    \includegraphics[width=\columnwidth]{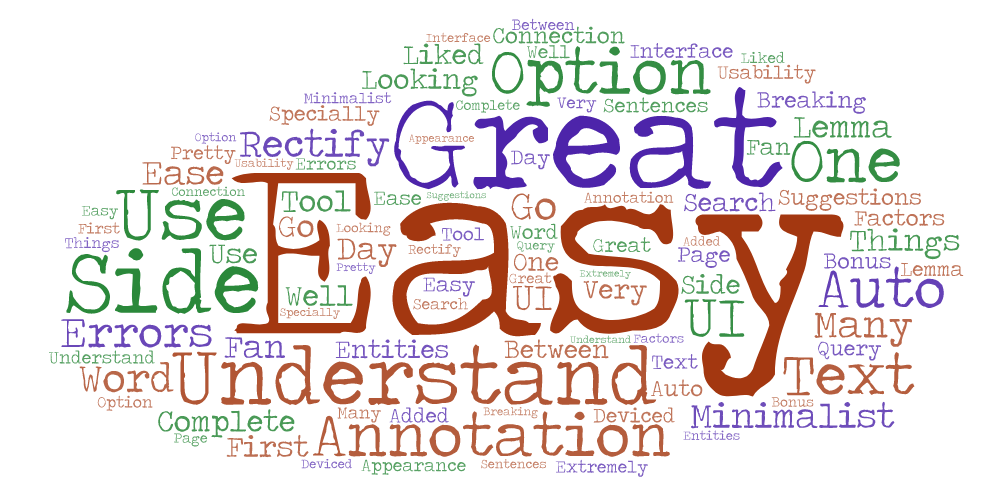}
	\vspace*{-4mm}
    \caption{Wordcloud of user testimonials}
    \Description{Wordcloud created using comments provided by users in the survey}
	\vspace*{-4mm}
    \label{fig:wordcloud}
\end{figure}

Table~\ref{table:surveyratings} shows the ratings given by the users.
Figure~\ref{fig:wordcloud} shows a word-cloud representation of the
testimonials provided by the users.

\subsection{Objective Evaluation}

Neves et al. \cite{neves2021extensive} used $26$ criteria classified in four
categories, viz., publication, technical, data and functional. We considered
four additional criteria A1, A2, A3 and A4. Criteria from publication category
(P1, P2 and P3) have not been considered.  Further, the criteria which were not
satisfied by any of the tools in the comparison (F2 and F6), have also been
omitted for the score calculation purpose.

Table~\ref{table:annoeval} lists the $25$ criteria that we have used for
evaluation of the annotation tools.

\begin{table*}[t]
    \begin{tabular}{llc|cccc|c}
        \toprule
        \multicolumn{3}{c|}{\bf Criteria} & \multicolumn{5}{c}{\bf Tools}\\
        ID & Description & Weight & WebAnno & doccano & FLAT & BRAT & Sangrahaka\\
        \midrule
        \vspace*{-3mm}
        \begin{filecontents*}{comparison.csv}
ID,Description,Weight,WebAnno,doccano,FLAT,BRAT,Sangrahaka
P1,Year of the last publication,0,1,0,0,1,1
P2,Citations on Google Scholar,0,1,0,0,1,0
P3,Citations for Corpus Development,0,1,0,0,1,0
T1,Date of the last version,1,1,1,1,0.5,1
T2,Availability of the source code,1,1,1,1,1,1
T3,Online availability for use,1,0,0,1,0,0
T4,Easiness of Installation,1,0,1,1,0.5,1
T5,Quality of the documentation,1,1,1,1,1,0.5
T6,Type of license,1,1,1,1,1,1
T7,Free of charge,1,1,1,1,1,1
D1,Format of the schema,1,1,1,1,0.5,1
D2,Input format for documents,1,1,0.5,1,1,1
D3,Output format for annotations,1,1,1,1,0.5,0
F1,Allowance of multi-label annotations,1,1,0,1,1,1
F2,Allowance of document level annotations,0,0,0,0,0,0
F3,Support for annotation of relationships,1,1,0,0,1,1
F4,Support for ontologies and terminologies,1,1,0,1,1,1
F5,Support for pre-annotations,1,0.5,0,0.5,0.5,0
F6,Integration with PubMed,0,0,0,0,0,0
F7,Suitability for full texts,1,0.5,0.5,1,1,1
F8,Allowance for saving documents partially,1,1,1,1,1,1
F9,Ability to highlight parts of the text,1,1,1,1,1,1
F10,Support for users and teams,1,0.5,0.5,1,0.5,0.5
F11,Support for inter-annotator agreement,1,1,0.5,0,0.5,0.5
F12,Data privacy,1,1,1,1,1,1
F13,Support for various languages,1,1,1,1,1,1
A1,Support for querying,1,0,0,0,0,1
A2,Server side crash tolerance,1,0,0,0,0,1
A3,Client side crash tolerance,1,0,0,0,0,1
A4,Web-based/distributed annotation,1,1,1,1,1,1
\end{filecontents*}
\csvreader[head to column names]{comparison.csv}{}{\\\ID&\Description&\Weight&\WebAnno&\doccano&\FLAT&\BRAT&\Sangrahaka}\\
        \midrule
        \multicolumn{2}{c}{\bf Total} & 25 & $18.5$ & $15.0$ & $19.5$ & $17.5$ & $\mathbf{20.5}$\\
        \multicolumn{2}{c}{\bf Score} & ~ & $0.74$ & $0.60$ & $0.78$ & $0.70$ & $\mathbf{0.82}$\\
        \bottomrule
    \end{tabular}
    \caption{Criteria for evaluation of annotation tools}
	\label{table:annoeval}
\end{table*}

\end{document}